\documentclass[12pt,titlepage]{article}
\usepackage{graphicx}
\usepackage{psfrag}
\usepackage{amssymb}
\usepackage{epsfig}
\usepackage{bm}
\font\grande=cmr10 scaled \magstep4
\font\medio=cmr10 scaled \magstep2
\outer\def\beginsection#1\par{\medbreak\bigskip
      \message{#1}\leftline{\bf#1}\nobreak\medskip
\vskip-\parskip
      \noindent}

\newcommand{\eq}{\begin{equation}}

\newcommand{\eqx}{\end{equation}}
\newcommand{\eqn}{\begin{eqnarray}}

\newcommand{\eqnx}{\end{eqnarray}}
\newcommand{\bi}{\begin{itemize}}

\newcommand{\ei}{\end{itemize}}

\newcommand{\ad}{{a^{\dagger}}}
\newcommand{\fd}{f^{\dagger}}

\newcommand{\Qd}{{Q^{\dagger}}}

\newcommand{\ra}{\rangle}
\newcommand{\la}{\langle}

\begin{document}
\titlepage

\begin{flushright}
\vspace{5mm}
CERN-PH-TH/2006-143\\
TPJU-09/2006
\end{flushright}
\vspace{15mm}
\begin{center}

\grande{ A supersymmetric
matrix model: \\
II. Exploring higher-fermion-number sectors}

\vspace{15mm}

\large{G. Veneziano}

\vspace{5 mm}

 {\sl Theory Division, CERN, CH-1211 Geneva 23, Switzerland }

{\sl and}

{\sl Coll\`ege de France, 11 place M. Berthelot, 75005 Paris, France}
\vspace{5 mm}

   \large{J. Wosiek}

   \vspace{5 mm}

   {\sl M. Smoluchowski Institute of Physics, Jagellonian University}

{\sl Reymonta 4, 30-059 Krak\'{o}w, Poland}

\end{center}
\vskip 10mm
\centerline{\medio  Abstract}
\vskip 4mm
\noindent
Continuing our previous analysis of a supersymmetric quantum-mechanical  matrix model,  we study in detail the properties of its  sectors with fermion number $F=2$ and $3$.
We confirm all
previous expectations, modulo the appearance, at strong coupling,  of {\it two} new bosonic ground states  causing a further
jump  in Witten's index  across  a previously identified critical
't Hooft coupling $\lambda_c$. We are able to elucidate the origin of these new SUSY vacua
by considering the
$\lambda \rightarrow \infty$ limit  and a strong coupling expansion around it.
\vspace{5mm}

\vfill
\begin{flushleft}
CERN-PH-TH/2006-143 \\
TPJU-09/2006\\
July  2006\\
\end{flushleft}

\newpage

\section{Introduction}

In a previous paper \cite{VW1} (see also \cite{Adriano} for motivations and details) we have introduced and discussed a  new Hamiltonian approach
to large-$N$ (planar) theories and have illustrated its effectiveness in a simple supersymmetric 
quantum-mechanics  model. It is
 the $N \rightarrow \infty$ limit of an $N \times N$ matrix model
defined by the following supersymmetric charges and 
Hamiltonian: 
\eq Q= {\rm Tr} [f \ad(1+g\ad)] =  {\rm Tr} [f A^{\dagger}],
\;\;\; \Qd=   {\rm Tr} [\fd (1+g a) a]=   {\rm Tr} [\fd A]\, . \eqx \eq
H=\{Q^{\dagger},Q\} = H_B+H_F  \, ,\label{h1} \eqx \eq H_B=  {\rm Tr}
[\ad a + g(\ad^2 a + \ad a^2) + g^2 \ad^2 a^2] \, ,\label{h2} \eqx
\eqn
H_F&=&  {\rm Tr}  [\fd f + g ( \fd f (\ad+a) + \fd (\ad+a) f) \nonumber \\
& + & g^2 ( \fd a f \ad + \fd a \ad f + \fd f \ad a + \fd \ad f a)] \, ,\label{h3}
\eqnx
where bosonic and fermionic destruction and creation operators satisfy:
\eq
[a_{ij},\ad_{kl}]=\delta_{il}\delta_{jk}  \label{com} \, , \, \, \{f_{ij} \fd_{kl}\}=\delta_{il}\delta_{jk} \, ,
\eqx
all other (anti)commutators being zero.

This dynamical  system turned out to be quite interesting per se. Since its
Hamiltonian conserves fermion number $F =  {\rm Tr}[\fd f ]$, it can be analysed
$F$-sector by $F$-sector, with SUSY connecting sectors differing
by one unit of $F$. The task of pairing states in supermultiplets is facilitated by
the introduction of another conserved operator \cite{VW1}: \eq C
\equiv [\Qd, Q]    \, , ~  [C, H] =0 \, , ~ C^2 = H^2 \, . \label{C}
\eqx Eigenstates of $H$ with eigenvalue $E>0$ can be classified
according to their ``$C$-parity", i.e. according to whether $C =
\pm E$. We may also consider the combination: \eq C_F = (-1)^F ~
\frac{C}{E} \, , \eqx a good quantum number for each  SUSY
doublet.  States with $C=+E ~(-E)$ are  annihilated by $\Qd~ (Q)$.
All the states  in the $F=0,1$ sectors turn out to have $C_F= -1$
but, in higher-$F$ sectors,  there also are some ``unnatural"
states  with $C_F = +1$:  these are important for the full matching
of bosons and fermions.

In \cite{VW1} we analysed in detail the $F=0$ and $F=1$ sectors of the model. They turn out to provide  a complete  (although highly reducible) representation of SUSY and to exhibit a number of interesting features. We briefly summarize them hereafter:
\begin{itemize}
\item
There is a (discontinuous) phase transition, as a function of the
't Hooft coupling $ \lambda \equiv g^2 N $, at $ \lambda =
\lambda_c =  1 $.  The mass/energy gap, which is present at $ \lambda < 1
$, disappears at $ \lambda = 1 $
 and reappears at $ \lambda >1 $.
\item
The Witten index \cite{WQM} (restricted to these two sectors) jumps by one unit at  $ \lambda = 1 $ as a result of the appearance, on top of the trivial Fock vacuum,
 of a new, normalizable zero-energy state at $ \lambda > 1 $. As a consequence, all higher
  supermultiplets rearrange at the transition point.
\item The system exhibits an  exact strong--weak duality in its spectrum, with  fixed point at $\lambda_c$: all excited eigenenergies at
$\lambda$ and $1/\lambda$  are connected by
a simple relation.
\item
The spectrum of the model can be computed analytically in terms of the zeros of a hypergeometric function.
This makes all the above properties explicit.
In particular, the details of the critical behaviour at $\lambda_c=1$ can be analysed.
\end{itemize}

In a subsequent paper \cite{OVW1}  some mathematical implications of the pairing of states at arbitrary
  $F$ and $B$ for the combinatorics of binary necklaces have been discussed.
Here
 we extend the analysis of \cite{VW1} to the $F=2$ and $F=3$ sectors.
As expected, unlike those with  $F=0$ and $F=1$, these two higher-$F$  sectors do {\it not} provide
a complete representation of SUSY (although they contain many of them). In other words, while all
$F=2$ states find their SUSY partner in the $F=3$ sector, the reverse is not true: some $F=3$ states
have no partner with  $F=2$; their partners  are  expected to lie, rather, in the $F=4$ sector.
The quantum number $C$, introduced above,  distinguishes the $F=3$ states that have SUSY partners with
$F=2$ from those with $F=4$
companions.
Thus the linear combinations of $F=3$ fermions that should be
degenerate with $F=2$ bosons can be neatly identified, at least for sufficiently weak coupling,
and these expectations can be compared with actual numerical calculations.

While we  find no surprises at weak coupling ($\lambda < 1$), the numerical spectrum at strong
coupling ($\lambda >1$) leads to something  unexpected: {\it two new} zero-energy bosonic states pop up,
 causing Witten's index to jump by two units at the $\lambda = 1$ phase transition.

The outline of the paper is as follows:
in the next section we construct the single-trace (planar) basis with $F=2,3$ and compare  it with
the  $F=0,1$ case discussed in  \cite{VW1,Adriano}.  We also derive explicit expressions for the leading-order  matrix elements of the Hamiltonian.
The increasing complexity (and some emerging regularities), as we move to higher fermionic sectors,
will be emphasized. Section 3 contains the detailed discussion of the physics based on
the numerical diagonalization of the planar Hamiltonian. Section 4 provides  an analytic construction
of the new susy vacua, first at infinite 't Hooft coupling, and then in the whole strong coupling phase via an expansion in $1/ \lambda$. We end by summarizing the main results of this work.

\section{$F = 2,3 $ Fock states and matrix elements at large $N$}
Since the Hamiltonian and SUSY charges are simple polynomials in creation and annihilation (c/a) operators,
it is advantageous to work in the gauge-invariant eigenbasis of the occupation number operators $B= {\rm Tr} [\ad a]$
and $F= {\rm Tr} [\fd f]$.
At infinite $N$, this basis substantially simplifies. Only  states created by single traces of fermionic and bosonic
operators give  leading-$N$ contributions. For this reason, the $F=0$ and $F=1$ Hilbert spaces were spanned by  basis vectors
labelled by just a single integer $n$ --the number of bosonic quanta. Extension to higher fermionic sectors is straightforward
but requires a little care. A generic state with two fermions has a form
\eqn
|n_1,n_2\ra = \frac{1}{{\cal N}_{n_1 n_2}} {\rm Tr} [\ad^{n_1}\fd\ad^{n_2}\fd]|0\ra ,\;\;\;n_1,n_2 \geq 0 \, ,
\eqnx
with a known normalization constant to be discussed shortly. Owing to the cyclic symmetry of the trace,
states differing by the interchange of $n_1$ and $n_2$ are linearly dependent (in fact $|n_1,n_2\ra = - |n_2,n_1\ra$
in this case). This fermionic minus sign has yet another consequence: there are no states with $n_1=n_2$. Therefore
the $F=2$ basis can be taken to be
\eqn
|n_1,n_2\ra,\;\;\; 0\leq n_1 < n_2 .
\eqnx

Similarly the basis states with three fermions are taken as
\eqn
|n_1,n_2,n_3\ra = \frac{1}{{\cal N}_{n_1 n_2 n_3}} {\rm Tr} [\ad^{n_1}\fd\ad^{n_2}\fd \ad^{n_3}\fd ]|0\ra,\;\;\;\;
 0\leq n_1 ,\;\;\;\;\ n_1 \leq n_2,\;\;\;\;\ n_1 \leq n_3.  \label{bf3}
\eqnx
Notice that cyclic symmetry does not imply any ordering of $n_2$ and $n_3$. Also, the cases where all (or some) of the bosonic occupation numbers coincide are not excluded in this sector.
All the above states are orthogonal in the planar limit, i.e. the off-diagonal elements of the inner-product matrix are subleading.

Matrix elements of the Hamiltonian (\ref{h1}--\ref{h3}) also simplify considerably at infinite $N$.
The ``planar rules"  for  obtaining  the leading contributions were formulated and illustrated in detail in
Refs.\cite{VW1} and \cite{Adriano}. In short: one uses  Wick's theorem, keeping only the colour contractions
that  give the maximal number of colour-index  loops, hence the highest power of $N$.
They come from the planar configurations of all c/a operators, as in the celebrated case of Feynman diagrams
\cite{'tH}.

The simplest application of the above rules is the calculation of the normalization factors. One obtains for $F=2$
\eqn
{\cal N}_{n_1 n_2} = \sqrt{N^{n_{{\rm tot}}}},
\eqnx
where $n_{tot}=n_1+n_2+2$ is the total number of quanta in a given state $|n_1,n_2\ra$.

There is more structure in the three-fermion sector:
\eqn
{\cal N}_{n_1 n_2 n_3} = \sqrt{d} \sqrt{N^{n_{{\rm tot}}}},\;\;\;n_{{\rm tot}}=n_1+n_2+n_3+3,
\eqnx
with the degeneracy factor
\eqn
d=\left\{ \begin{array}{cc}
      3 & {\rm if} \;\; n_1=n_2=n_3, \\
      1 & {\rm otherwise}.
      \end{array}
      \right.
\eqnx

We are now ready to calculate the complete matrix elements of various terms of the Hamiltonian (\ref{h1}--\ref{h2}).
The algebra is straightforward, though a little tedious. We stress that the above normalization
factors are crucial. 
Our final result  for the $F=2$ sector reads:
\eqn
\langle n_1,n_2|H|n_1,n_2\rangle = (n_1+n_2+2)(1+b^2) - b^2 (2 - \delta_{n_1,0}) - 2 b^2 \delta_{n_2,n_1+1}, \\
\langle n_1+1,n_2|H|n_1,n_2\rangle = b (n_1+2) = \langle n_1,n_2|H|n_1+1,n_2\rangle,\\
\langle n_1,n_2+1|H|n_1,n_2\rangle = b (n_2+2) = \langle n_1,n_2|H|n_1,n_2+1\rangle.\\
\langle n_1+1,n_2-1|H|n_1,n_2\rangle  = \langle n_1,n_2|H|n_1+1,n_2-1\rangle = 2 b^2 (1-
 \delta_{n_2,n_1+1}) \, , \label{last}
\eqnx where, as in  \cite{VW1}, we have introduced, for
convenience, $b \equiv \sqrt{\lambda}$. Note the
$\delta_{n_2,n_1+1}$ exception in the first and last equations.
For this configuration of the bosonic occupation numbers 
Eq. (\ref{last}) gives rise to a diagonal matrix element. The
$\delta_{n_2,n_1+1}$ function prevents double counting of this
contribution. Notice also that it comes with the opposite sign
to that suggested in Eq. (\ref{last}). This is because the resulting
state $\langle n_1+1,n_2-1|$  equals  minus the state $\langle
n_2-1,n_1+1|$, which belongs to our basis.

In the $F=3$ sector we obtain:
\eqn
\langle n_1,n_2, n_3|H|n_1,n_2,n_3\rangle = (n_1+n_2+n_3 +3)(1+b^2) 
- b^2 (3 - \delta_{n_1,0} -
\delta_{n_2,0}- \delta_{n_3,0})
\label{diag3} , \\
\langle n_1+1,n_2,n_3|H|n_1,n_2,n_3\rangle = \langle n_1,n_2,n_3|H|n_1+1,n_2,n_3\rangle = b (n_1+2) \Delta
\label{off31} \, , \\
\langle n_1+1,n_2-1,n_3|H|n_1,n_2, n_3\rangle  =  \langle n_1,n_2,n_3|H|n_1+1,n_2-1, n_3\rangle b^2 \Delta
\label{off32} \, ,
\eqnx
as well as cyclic permutations of (\ref{off31}), (\ref{off32}), where $\Delta=1/\sqrt{3}$ if $n_1=n_2=n_3$ and $\Delta=\sqrt{3}$ if the final state is of this form;
otherwise $\Delta=1$.
Similarly to the $F=2$ case, for some sets of $n_i$, Eq. (\ref{off32}) gives rise to diagonal
elements. However, the $F=3$ Hilbert space is considerably richer that the $F=2$ one and many other ``degenerate
cases" occur here as well.
Instead of dealing explicitly with all these ``exceptional" configurations, we adopt a simple, unifying rule:
\begin{center}
\begin{minipage}{12cm}
All contributions listed in (\ref{diag3}--\ref{off32}) should be added to the appropriate locations
in the $H_{IJ}$ table ($I$ and $J$ are the linear, composite indices labelling all states satisfying
(\ref{bf3})).
\end{minipage}
\end{center}

This rule also covers other ``degenerate" situations. For example,
for $n_1=n_2=n_3$ the ``plus cyclic" qualifier implies that the same matrix element receives an additional
factor 3. According to our rule, however, the same elementary matrix element should be added three times, which is of course equivalent.

\section{Results}

As for  the  $F=0$ and $F=1$ cases, the complete spectrum with two and three fermions was obtained by diagonalizing numerically the
above-mentioned  Hamiltonian matrices. To this end, we have introduced a cutoff $B_{{\rm max}}$ that limits the total number of
bosonic quanta in the system:
\eqn
B \leq B_{{\rm max}} \, .
\eqnx
Such a cutoff was found very useful in many applications \cite{JW}, \cite{CW}, since a) it can be easily implemented
in our bases, b) it preserves many symmetries in more complex models, and c) the spectra converge
 well with increasing $B_{{\rm max}}$.

\subsection{Dynamical supermultiplets}
The situation is not different in the present case. At $\lambda=2.0$, for example, the lower
(i.e. first 20) levels have converged to five decimal
digits for cutoffs $B_{{\rm max}}=40~(30) $ for $F=2~ (3) $. Figure 1 compares the spectra
in all four fermionic sectors, displaying clearly similarities and differences between the $F=0, 1$ and the
 $F=2, 3$ cases. Supersymmetry, which is broken by the cutoff, is nicely restored -- there is an excellent
 boson fermion degeneracy already at the above or higher values of $B_{{\rm max}}$. Interestingly, while all $F=0$ states had their
 supersymmetric counterparts in the $F=1$ sector and {\em vice versa}, for higher fermionic numbers this is not so.
 Every state with  $F=2$ has its partner in the $F=3$ sector, but the reverse is not true:  some states with  $F=3$ are ``unpaired" in Fig. 1 and, consequently, must have their counterparts in the $F=4$ sector. This is already expected
 at  the level of counting basis states in respective Hilbert spaces;  however, Fig. 1 provides a clear dynamical
 confirmation of this structure.

\begin{figure}[tbp]
\psfrag{yyy}{${\scriptstyle E}$ }\psfrag{ttt}{${\scriptstyle\lambda=2.0}$}
\psfrag{xx1}{${\scriptstyle F=0\hspace*{2.5cm}F=1}$}\psfrag{tt1}{${\scriptstyle\lambda=0.5}$}
\psfrag{fff}{${\scriptstyle F=2\hspace*{2.5cm}F=3}$}
\epsfig{width=8cm,file=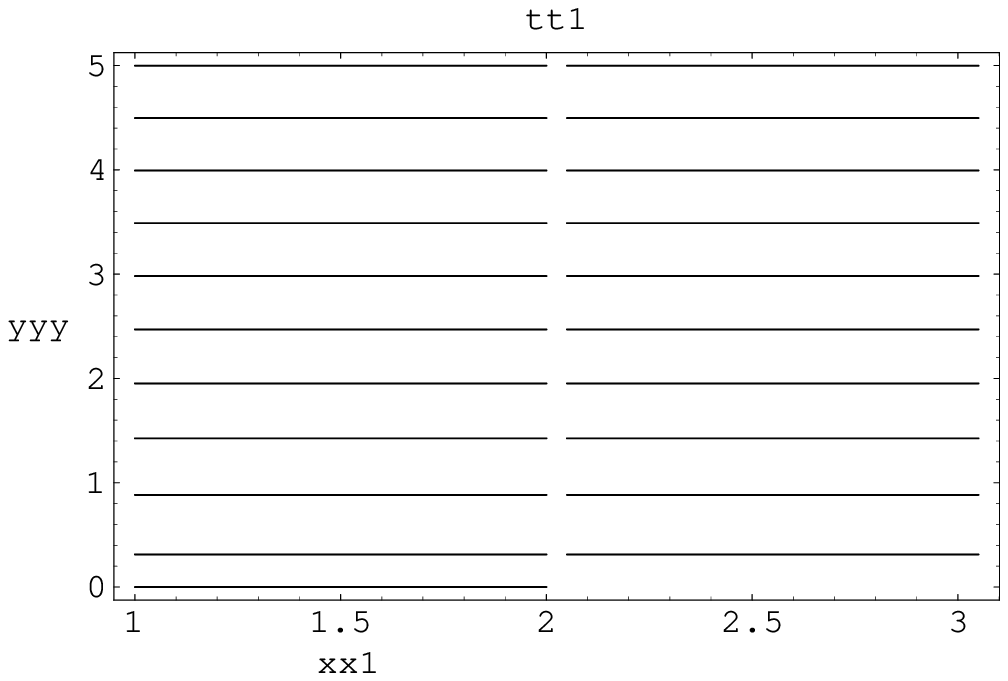}
\epsfig{width=8cm,file=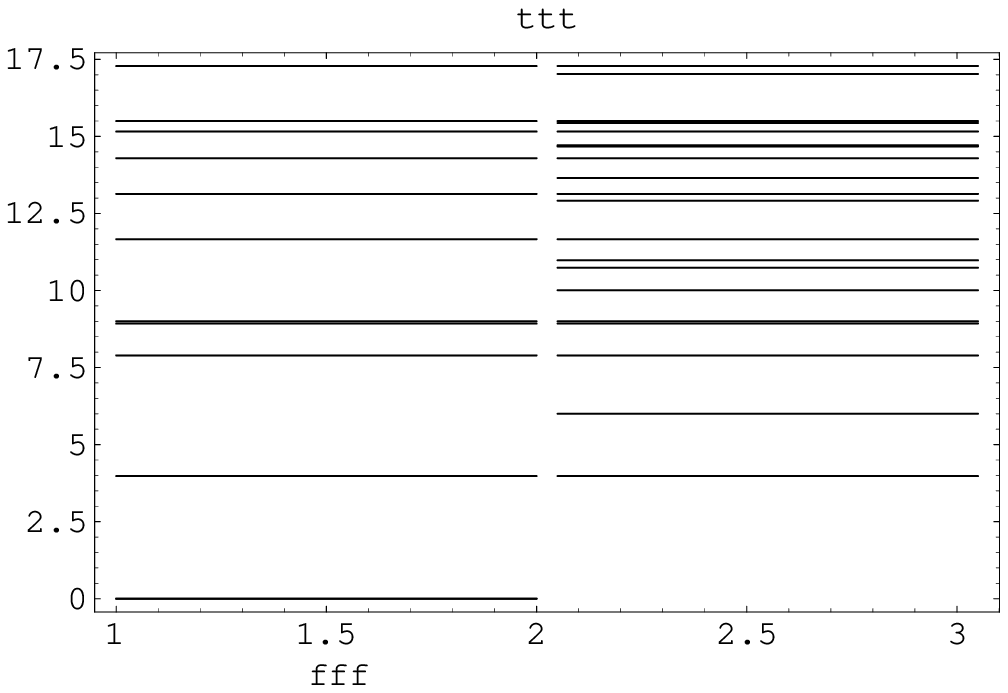}

\vskip-4mm \caption{Low-lying bosonic and fermionic levels in the
first four fermionic sectors.} \label{fig:susyf01}
\end{figure}

Apart from supersymmetry, the spectra in higher-$F$ channels are obviously much richer than
 those for $F=0,1$. In fact in all earlier cases we know of, the energy levels were almost
 equidistant in the  large-$N$ limit \cite{MP},\cite{MO}. This also happened for  $F=0,1$ in our model, possibly suggesting
  some generic simplification of planar dynamics, but is no longer true with more fermions. In fact, some
 levels are so close that they seem to merge into one because of the poor resolution of the graph
 (e.g. the levels \# 4,5 with $F=2$ and \# 4,5 with $F=3$\footnote{This splitting has been clearly established and is
 {\em not} a finite cutoff effect.}).  Interestingly, both members of the $\# 4, 5$, $F=2$ doublet find their
 partners  within the $F=3$ sector.  However, this is by no means a general rule (cf. \# 10,11 with $F=3$ ).

\subsection{Phase transition, rearrangement and new SUSY vacua}
One of the remarkable features of our system, seen also
analytically in the $F=0,1$ sectors, was the phase transition at
$\lambda = \lambda_c=1$. At this point the spectrum became
gapless, and a critical slowing down of the cutoff dependence was
observed. The same phenomenon occurs in the $F=2$ and $F=3$
sectors. In Fig. 2 we show the $\lambda$ dependence of the few
lowest levels, with two and three gluinos. The critical slowing
down of the convergence is clearly observed. Close to the critical
point, supersymmetry is visibly broken, at fixed $B_{{\rm max}}$, while
away from $\lambda=1$ supermultiplets are well formed.

Interestingly, SUSY partners rearrange themselves across the phase-transition point,
as in the $F=0,1$ case.  A novel feature, however, is that, in the strong coupling phase,  {\it two} zero-energy states appear in the $F=2$ sector. This is to be compared with the $F=0$ case where the  trivial Fock vacuum (present at all
$\lambda$) is  accompanied by a second $E=0$ state in the strong coupling phase.

The claim that  all the curves in Fig. 2 converge  to zero at
$\lambda=1$ may seem a little premature.  However, we have repeated
our  analysis for a few values of $B_{{\rm max}}$ and established that
this is the most likely scenario. This point is now being carefully
and quantitatively studied  by P. Korcyl. The analytic results
discussed in the next section also confirm this conclusion.

\begin{figure}[tbp]
\psfrag{xxx}{${\scriptstyle\sqrt{\lambda}}$} \psfrag{yyy}{${\scriptstyle E}$ }
\psfrag{ttt}{${\scriptstyle F=2\;\; {\rm and}\;\; F=3}$}
\begin{center}
\epsfig{width=8cm,file=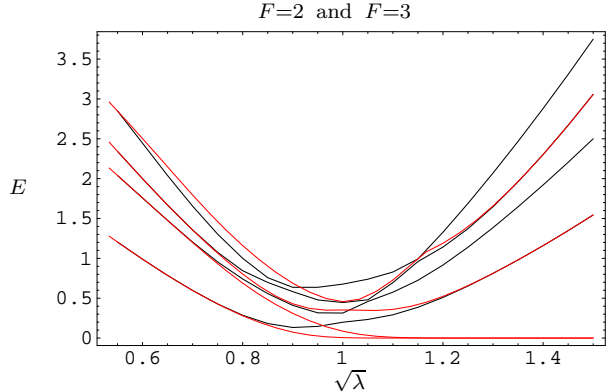}
\end{center}
\vskip-4mm
\caption{Rearrangement of the $F=2$ (red) and $F=3$ (black) levels while passing through the critical
coupling $\lambda_c=1$; $B_{{\rm max}} = 30~ (25)$ for $F=2 ~(3)$.}
\label{fig:cfplot}
\end{figure}

\subsection{Supersymmetry fractions}

Apart from the unbalanced vacua and boson--fermion degeneracies of higher states, supersymmetry manifests itself
in yet a different but important way. Namely, the supersymmetry charges $Q$ and $\Qd$ transform members of the supermultiplets
into each other. Since our method provides not only the eigenenergies, but also the eigenstates, we can directly
check these relations. To this end define the ``supersymmetry fractions" \cite{CW}
\eqn
q_{mn}=\sqrt{\frac{2}{E_n+E_m}}\la F=3,E_m|\Qd|F=2,E_n\ra,
\eqnx
which are the, suitably normalized,  coordinates of the supersymmetric image of  the $n$-th, $F=2$, eigenstate in the $F=3$ sector.
For unbroken supersymmetry, $q_{mn}=\delta_{mn}$ (with an appropriate labelling of the $F=3$ states). For finite cutoff, however,
the supermultiplets are not yet well formed and $q's$ are not unity \footnote{Even with  unbroken supersymmetry,
supersymmetry fractions may not be exactly 1. This happens when there is an exact degeneracy {\em between supermultiplets}.
In such a situation $q's$ measure directly the mixing angles of the members of supermultiplets.}.
Figure 3 shows the cutoff dependence of the first few supersymmetry fractions. For low cutoffs they can vary rather irregularly until  the partner is rapidly found  and  convergence takes place  up to a small noise.

\begin{figure}[tbp]
\psfrag{yyy}{$q_n+n-1$}\psfrag{xxx}{$B_{max}$}
\begin{center}
\epsfig{width=8cm,file=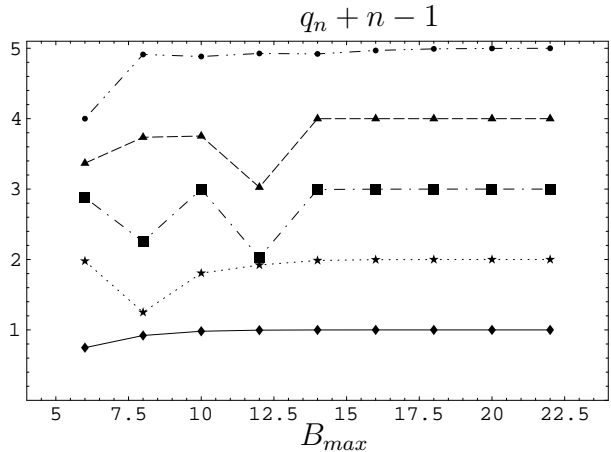}
\end{center}
\vskip-4mm \caption{First five supersymmetry fractions, defined as
$q_n={\rm Max}_m(|q_{mn}|)$, for a range of cutoffs.} \label{fig:susyf01}
\end{figure}

\subsection{Restricted Witten index}

The effect of the new SUSY vacua on the Witten index is best seen if we restrict
the sum
\eqn
W(T,\lambda)= \sum_i (-1)^{F_i} e^{-T E_i}
\eqnx
to the states in the $F=2$ and $F=3$ sectors and exclude the $C=-E, F=3$ states. In practice we have used
the supersymmetry fractions and identified the SUSY partners for each of the $F=2$ states.
This procedure, performed necessarily at finite cutoff, is unambiguous away from the transition point. Close to
$\lambda_c$, however,  supersymmetry is badly broken at any finite $B_{{\rm max}}$, and a definition of  what is meant by the
``energy of the supersymmetric partner" must be given. We made the following choice:
\eqn
W_R(T,\lambda)= \sum_i  \left( e^{-T E_i} -  e^{-T \bar{E}_i} \right),\;\;\;
\bar{E}_i=\frac{\sum_f E_f |q_{fi}|^2}{\sum_f |q_{fi}|^2} \, ,
\eqnx
where the $i(f)$ indices run over all states in the $F=2~(3)$ sectors. In another words, we take for the
energy of the partner the weighted average over all $F=3$ states with the weight given by the
supersymmetric fractions.
Away from $\lambda_c$ this selects the true supersymmetric partners and
automatically excludes  the $C=-E, F=3$ states.
The index defined in this way is a smooth function of the 't Hooft coupling (see Fig. 4) showing
the onset of the discontinuity caused by the new vacua in the strong coupling phase.

\begin{figure}[tbp]
\psfrag{yyy}{${\scriptstyle I_W(6)}$}\psfrag{xxx}{${\scriptstyle \sqrt{\lambda}}$}
\begin{center}
\epsfig{width=8cm,file=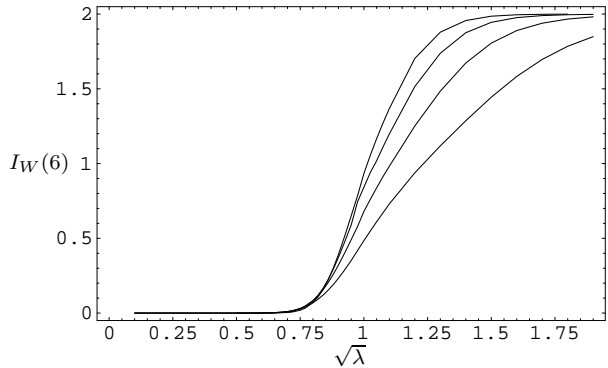}
\end{center}
\vskip-4mm \caption{Behaviour of the restricted Witten index, at $T=6$, around the phase transition.
The various curves correspond to $8 \le B_{{\rm max}} \le 14 $.} \label{fig:susyf01}
\end{figure}

\section{Explicit construction of the strong-coupling vacua}
The (numerical)  appearance of two new SUSY vacua above $ \lambda =1$ was quite unexpected.
In order to understand their origin we first consider the limit of infinite  't Hooft coupling.
As we will  see,  the necessity of  two new vacua comes out
very simply at  $ \lambda = \infty$. The corresponding eigenvectors also greatly simplify in that limit
(they have projections only on
 one or two Fock states).
 We will then  show, by an explicit construction,  that those  vacua
persist even at finite (but sufficiently large)   $ \lambda$, although the corresponding eigenvectors will contain an infinite superposition of Fock states.

We wish to mention that the  $ \lambda \rightarrow \infty$ limit also simplifies the discussion of our model at arbitrary $F$, since the Hamiltonian becomes block-diagonal with finite-dimensional blocks.
Furthermore, some restricted supertraces can be used to identify (and/or guess) the blocks
where  new SUSY vacua are expected. This result was exploited in Ref. \cite{OVW1} and will be
described in more detail elsewhewe.

\subsection{SUSY vacua at infinite $ \lambda$}

The limit $ \lambda  \rightarrow \infty$ of our Hamiltonian is actually infinite. However, we can  define
appropriately rescaled SUSY charges that have a finite limit and that, consequently, define a finite rescaled Hamiltonian. Of course eigenvectors will be invariant under such an overall rescaling.
The strong-coupling limit of our rescaled SUSY charges reads simply:
\eqn
 Q_{\rm{strong}}=\lim_{\lambda\rightarrow\infty} \frac{1}{\sqrt{\lambda}} Q =\frac{1}{\sqrt{N}} {\rm Tr} (\ad^2 f),\;\;\;\;
Q_{\rm{strong}}^{\dagger}=\lim_{\lambda\rightarrow\infty} \frac{1}{\sqrt{\lambda}} \Qd =\frac{1}{\sqrt{N}} {\rm Tr} (a^2 \fd).
\eqnx

The peculiar property of these charges is that they change $F$ and $B$ but preserve the
combination $2F+B$ (unlike the zero-coupling charges that preserve $F+B$).
In other words, if we represent our base states on a grid having coordinates $F$ and $F+B$, the
 infinite $ \lambda$ charges connect states along a 45-degree diagonal (while the zero-coupling charges connect  states on lines parallel to the $F$ axis).
 A necessary condition for having no zero-energy states along these diagonals is that the corresponding supertrace be zero. Conversely, if the supertrace is non-zero, there must be some unpaired SUSY vacua lying on those diagonals.

 A very simple inspection immediately shows that the latter is the case for the diagonal that contains
 the state with $F=0$ and $B=1$ (the strong coupling vacuum of \cite{VW1}) as well as for those that contain either the $F=2, B=1$ or the  $F=2, B=3$ states.
 The latter are our new SUSY vacua at  infinite $ \lambda$.

 Let us now consider the  infinite $ \lambda$ limit of the (rescaled)  Hamiltonian (\ref{h1})--(\ref{h3}):
\eqn
 H_{\rm{strong}}=\lim_{\lambda\rightarrow\infty} \frac{1}{\lambda} H =
 \frac{1}{N}{\rm Tr} [ \ad^2 a^2 + \fd a f \ad + \fd a \ad f + \fd f \ad a + \fd \ad f a] \, .
\eqnx
 This result can be simplified further,  providing a  good illustration
 of our planar rules.
Namely, the third and  fourth terms must be brought to the normal form. Explicitly
\eqn
\fd_{ij} a_{jk} \ad_{kl} f_{li} &=& \fd_{ij}(\ad_{kl}a_{jk} + \delta_{jl}\delta_{kk}) f_{li}=N {\rm Tr} [\fd f],\\
\fd_{ij} f_{jk} \ad_{kl} a_{li} &=& \fd_{ij} \ad_{kl} f_{jk}  a_{li} = 0 \, ,
\eqnx
since the neglected terms do not yield a single trace.
The strong-coupling Hamiltonian thus reads
\eqn
H_{{\rm SC}}= \fd f +
 \frac{1}{N}{\rm Tr} [\ad^2 a^2 + \ad \fd a f + \fd\ad f a] \, .  \label{hstr}
\eqnx
Remarkably, it conserves {\em both} $F$ {\em and} $B$.

\subsection{ Strong-coupling vacua from infinite-coupling vacua}

In the $F=0$ sector we were able to give an explicit construction of the non-trivial vacuum  at $\lambda >1$. Actually, the new zero-energy state could be written down for any value of
$\lambda$ but was only normalizable in the strong-coupling region. We may ask whether a similar
analysis can be  carried out in the $F=2$ sector.  The $F=0$ and  $F=2$ sectors share the property that all their states are annihilated by $Q$. Therefore, null states  with  $F=0,2$ are characterized by the fact that they are  annihilated by $\Qd$. Indeed, the $F=0$ null state can  be easily obtained in this way by writing:
\eqn
Q &=& Q_w + Q_s \, , \, ~ \Qd = \Qd_w + \Qd_s  \, , \, \nonumber \\
Q_w &=&  {\rm Tr}  [f \ad]  \, , \,  ~ Q_s = g {\rm Tr}  [f \ad^2]
\;\;\; \Qd_w=  {\rm Tr}  [\fd  a]  \; , \;\; \Qd_s =  g {\rm Tr} [\fd a^2] \, ,
\eqnx
where the labels $w$, $s$ refer to the weak and strong coupling forms of the supersymmetric charges, respectively. Note that $\{Q_w,Q_s\} =0$ and that, if we define $H_w = \{Q_w,\Qd_w\}$, $H_s = \{Q_s,\Qd_s\}$, we also have $[H_w, Q_w] = [H_s, Q_s]=0 $, and similarly with $Q \rightarrow \Qd$.

Given the action of  $\Qd_w$ and $\Qd_s$ on  $F=0$ states:
\eq
\Qd_w |n \rangle = \sqrt{n} |n-1,F=1 \rangle \, , \, ~ \Qd_s |n \rangle = b \sqrt{n} |n-2, F=1 \rangle \, ,
\eqx
we see immediately that:
\eq
|v, F=0 \rangle  = \sum_{n=1}^{\infty}\frac{ ( -b)^{-n}}{\sqrt{n}}   |n \rangle
\eqx
is annihilated by $\Qd$. Clearly, it is normalizable only for $b >1$.

A very similar construction works  for one of the two strong-coupling ground states.
Let us look indeed for a state consisting of a linear superposition of $F=2$ states of the form:
\eq
|v, F=2\rangle _1 = \sum_{n=1}^{\infty}  c_n  |0, n \rangle \, .
\eqx
This time we find:
\eq
\Qd_s |0,1 \rangle =0 ~,~  \Qd_w |0, n \rangle = b^{-1} \Qd_s|0, n+1 \rangle \, ,
\eqx
and therefore a zero-energy state is simply:
\eq
|v, F=2\rangle _1   = \sum_{n=1}^{\infty} ( -b)^{-n}  |0, n \rangle \, .
\eqx
Note that the above state is written as a Taylor series in
$1/b$ starting from one of the two infinite-coupling ground states:
\eq
|v, F=2\rangle _1^{\infty} = |0,1 \rangle \, .
\eqx
Like its analogue in the $F=0$ sector,  this null state is only normalizable  if $b>1$.

For a second independent $E=0$ state,  let us start our series from the other infinite-coupling vacuum:
\eq
 |v, F=2\rangle _2^{\infty}\equiv|0,3\rangle - 2 |1,2\rangle \rightarrow \, |v, F=2\rangle _2
\label{SCnull}
\eqx
and use the following theorem:

\begin{center}
\begin{minipage}{12cm}
 Given an $F=2$ state annihilated by $\Qd_s$ (e.g.  (\ref{SCnull})) the state:
\eq
|v, F=2\rangle _2 \equiv \left(1 +  Q_s H_s^{-1} \Qd_w\right)^{-1} |v, F=2\rangle _2^{\infty}
\eqx
is annihilated by $\Qd$ and is therefore a null state.
\end{minipage}
\end{center}

The proof is easily carried out by expanding the fraction in powers of $1/b$ (NB: $Q_s H_s^{-1} = O(1/b) $)  and by noting that the first term is annihilated (by construction)  by $\Qd_s$ while the other terms cancel pairwise.  Indeed:
\eqn
&&  \Qd_s \left(Q_s H_s^{-1} \Qd_w\right)^{n +1} = ( \Qd_w  + H_s^{-1}Q_s \Qd_w \Qd_s ) \left(Q_s H_s^{-1} \Qd_w\right)^n \nonumber \\ &=& \Qd_w  \left(Q_s H_s^{-1} \Qd_w\right)^n +
H_s^{-1}Q_s \Qd_w Q_s H_s^{-1} \Qd_w \Qd_s \left(Q_s H_s^{-1} \Qd_w\right)^{n-1}\nonumber \\
&=& \dots =
 \Qd_w  \left(Q_s H_s^{-1} \Qd_w\right)^n \, ,
\eqnx
where we have used $\Qd_w^2 =0$ and have  iterated the procedure until $\Qd_s$ annihilates the state (\ref{SCnull}).

The above construction  can be generalized to the case in which the states in the chosen sector are not
necessarily annihilated by $Q_w$ and/or $Q_s$. This is relevant in sectors with $F>2$.
The claimed generalization is that the following state is annihilated by $\Qd$ {\it and } $Q$:
\eq
|v \rangle  \equiv \left(1 +  Q_s H_s^{-1} \Qd_w + \Qd_s H_s^{-1} Q_w \right)^{-1} |v \rangle^{\infty}\,.
\eqx

 The proof (left as an excercise) is facilitated by the following two identities:
 \eqn
 \Qd_w \Qd_s  \left( Q_s H_s^{-1} \Qd_w + \Qd_s H_s^{-1} Q_w \right)^n &=& \Qd_w  \left( Q_s H_s^{-1} \Qd_w  \right)^n \Qd_s  \nonumber \\
 Q_w Q_s  \left( Q_s H_s^{-1} \Qd_w + \Qd_s H_s^{-1} Q_w \right)^n &=& Q_w  \left( \Qd_s H_s^{-1} Q_w  \right)^n Q_s \, .
 \eqnx

\section{Summary}

In this paper we have extended our previous work  on the $F =0, 1$ sectors  of a supersymmetric matrix model \cite{VW1}  to sectors with fermion number $F=2, 3$.
 In Section 1 we have summarized our previous results:  hereafter we will do the same for the two new sectors, underlining similarities and differences.

\bi

\item   As in the $F=0,1$ case,  here too  there is a phase transition at  $\lambda = \lambda_c=1$: while the spectrum is discrete below and above $\lambda_c$, the energy-gap
disappears (the spectrum becomes continuous) exactly at
$\lambda_c$.
 There is also, in both cases, a critical slowing down of convergence (as a function of
 the cutoff $B_{{\rm max}}$)  in the vicinity of $\lambda_c$.

\item Again in analogy with  what was found in the $F=0,1$ sectors, supersymmetry, which is broken by the cutoff, is quickly restored for
$\lambda\ne\lambda_c$,  giving  exact degeneracy of fermionic and bosonic
eigenstates. This degeneracy is indeed SUSY-driven:  with the aid of suitably defined
``supersymmetry fractions", we have verified that the degenerate eigenstates are supersymmetric
images of each other.

\item While in the low-coupling phase there are no zero-energy eigenstates in these sectors,
{\it two} such states appear for $\lambda > \lambda_c$ in the  $F=2$ sector (with a consequent jump of Witten's index by two units). This is similar (but not identical)
 to the $F=0,1$ case, where the empty Fock state (a zero-energy eigenstate for all
 values of $\lambda$) is accompanied  by another, non-trivial SUSY vacuum, just  in the
 strong-coupling phase. Such a  ``popping up" of new SUSY vacua is only possible,  at finite cutoff,
thanks to explicit supersymmetry breaking around $\lambda_c$.
 In fact our numerical results
 reveal, as in the $F=0,1$ case, the rearrangement of members of all supermultiplets while crossing
 the transition point at any finite value of  $B_{{\rm max}}$.
 At infinite cutoff (i.e.  in complete absence of SUSY breaking) this rearrangement is made possible by the disappearance of the energy gap at $\lambda_c$.

\item A new feature of multifermion states is the {\em intertwining}
of the $F= 2, 3$ and $F=3, 4$ supermultiplets: while all $F=2$ states have their partners
in the $F=3$ sector, some $F=3$ states remain unpaired: they must form new supersymmetry doublets
with the $F=4$ states. This behaviour  must repeat itself {\em ad infinitum} in $F$ since
the sizes of bases, at fixed $B$, are monotonically growing with $F$.
This was not the case for $F=0,1$, where all doublets were complete.
In other words, beginning with three ``gluinos", eigenstates with given $F$
may (and will) carry both signs of the $C$-parity quantum number (\ref{C})  introduced in \cite{VW1}.

\item This last point implies that defining the Witten index for the $F=2,3$ sectors
requires more care than in the $F=0,1$ case. Namely, one should sum only over
complete supermultiplets and not over all $F=2$ and $F=3$ states. 
This was done using once more
the supersymmetry fractions and the resulting object does indeed exhibit a discontinuity
at $\lambda=\lambda_c$, corresponding to the number of new vacua appearing across $\lambda_c$. 

\item  We did not find, in these higher-$F$ sectors,  any evidence for the  strong-weak duality
discovered in \cite{VW1}.

\item  Unlike in the $F=0,1$ sectors, we were not able to solve analytically for the spectrum.
Indeed,  the structure of the spectra with $F=2,3$  is much more
complex than the one with $F=0,1$. In particular, the eigenenergies are no longer approximately equidistant,
as was   the case for $F=0,1$ and  for other matrix models discussed
in the literature. Thus approximate equidistance  cannot be considered as a generic property of infinite $N$ dynamics.

\item
Finally, by considering the $\lambda \rightarrow \infty$ limit of our model and a strong-coupling expansion around it, we have understood the origin
of (and analytically constructed)  the new  bosonic zero-energy states at strong coupling.
The argument shows that the occurrence of such bosonic vacua should extend to arbitrary values
of $F$.
In the infinite-coupling limit the (rescaled) Hamiltonian becomes block-diagonal with finite-size
blocks characterized by a fixed number of fermions {\it and} bosons. This limit  has interesting
implications for the combinatorics of binary necklaces \cite{OVW1} and for the
spectrum of (non-supersymmetric) spin-chain models, as explained in a forthcoming paper.
\ei

\section*{Acknowledgements}
We would like to thank E. Onofri for useful discussions. This work is partially supported by the
grant No. P03B 024 27
(2004 - 2007) of the Polish Ministry of Education and Science.


\begin{thebibliography}{99}
\bibitem{VW1} G. Veneziano and J. Wosiek, JHEP  {\bf 0601}  (2006) 156  \ [hep-th/0512301].
\bibitem{Adriano} G. Veneziano and J. Wosiek, to appear in Adriano Di Giacomo's Festschrift (2006)\ [hep-th/0603045].
\bibitem{WQM} E. Witten, Nucl. Phys. {\bf B185} (1981) 513; {\bf B202} (1983) 253.
\bibitem{OVW1} E. Onofri, G. Veneziano and J. Wosiek, {\em Supersymmetry and Combinatorics},  \ [math-ph/0603082].
\bibitem{'tH} G. 't Hooft, Nucl. Phys. {\bf B72} (1974) 461; see also G. Veneziano,
Nucl. Phys. {\bf B117} (1976) 519.
\bibitem{JW} J. Wosiek, Nucl.\ Phys. {\bf B644 } (2002) 85 \ [hep-th/0203116].
\bibitem{CW} M. Campostrini and J. Wosiek, Nucl.\ Phys. {\bf B703 } (2004) 454 \ [hep-th/0407021].
\bibitem{MP} E. Marinari and G. Parisi, Phys. Lett. {\bf B240} (1990) 375.
\bibitem{MO} G. Marchesini and E. Onofri, Phys. Lett. {\bf B240} (1990) 375.
\end{thebibliography}
\end{document}